
\input harvmac.tex

\def\exp{{\rm exp}}

\def\frac#1#2{{#1\over#2}}

\Title{\vbox{\baselineskip12pt\hbox{CLNS-95/1360}
                \hbox{hep-th/9509037}}}
{\vbox{\centerline{
A note on the deformed Virasoro algebra}}}

\centerline{Sergei Lukyanov \footnote{$\dagger$}
{On leave of absence from L.D. Landau Institute for Theoretical
Physics, Kosygina 2, Moscow, Russia}
\footnote{$^*$}{e-mail address: sergei@hepth.cornell.edu}}
\centerline{Newman Laboratory,}
\centerline{Cornell University, Ithaca, NY,  14853-5001, USA}
\centerline{}
\centerline{}

\centerline{\bf{Abstract}}
A current
of the  deformed Virasoro algebra
is identified with the Zamolodchikov-Faddeev operator for
the basic scalar particle in the  XYZ model.
\Date{September, 95}

\eject
The studies  of infinite dimensional algebras
has turned out to be  the  main tendency   in the
recent
development of the theory
of  two-dimensional quantum  integrable systems.
This approach was originated by the fundamental  work of A.A. Belavin,
A.M. Polyakov and A.B.  Zamolodchikov. Now it has become clear
that the  Virasoro-type  symmetry is not a unique
feature of Conformal Field Theory  models.
In the work\ \ref\lp{Lukyanov, S.  and   Pugai, Ya.: Bosonization
of ZF algebras: Direction toward deformed Virasoro
algebra.
Rutgers preprint RU-94-41 (1994) (hep-th/9412229)}\   an existence
of a   nontrivial deformation of the Virasoro algebra was
conjectured (see also \
\ref\fre{Frenkel, E. and Reshetikhin, N.: Quantum affine algebras
and deformations of the Virasoro and \ $W$-algebras. Preprint (1995)
(q-alg/9505025)} ).
It is expected to be
an algebra of dynamical symmetry in the XYZ
Heisenberg chain with the Hamiltonian:
\eqn\xyz{H=- \sum_n(J_x \sigma_n^x\otimes\sigma_{n+1}^x+
J_y\sigma_n^y\otimes\sigma_{n+1}^y+
J_z\sigma_n^z\otimes\sigma_{n+1}^z)\ ,}
where
$$ J_x>J_y> | J_z|\  .$$
Precisely speaking, the  symmetry with respect  to
the deformed Virasoro algebra (DVA)
arises  in
the thermodynamic limit only,
when the number of sites  in the chain goes to
infinity.
DVA  plays the same role as\ $ U_q(\widehat{sl(2)})$
at the level one for the
anti-ferromagnetic XXZ chain \ $(-J_z>J_x=J_y>0)$\
\ref\jde{Jimbo, J. and Miwa, T.:
Algebraic analysis of Solvable Lattice
Models, Kyoto Univ., RIMS-981 (1994)}.
The  bosonization  procedure of\ \lp\  allowed one to  construct
irreducible  representations and intertwining vertex operators for DVA
without using an explicit form
of its  commutation relations.
In the recent important paper \ref\yap{Shiraishi, J., Kubo, H., Awata,
 H. and Odake, S.: A quantum deformation of the
Virasoro algebra and the
Macdonald symmetric functions. Preprint YITP/U-95-30, DPSU-95-5,
UT-715 (1995) (q-alg/9507034)}
the generators of DVA  were constructed explicitly
by
the same  bosonization procedure.
This
construction was already generalized for
the\ $ WA_n$\  algebras \ \ref\ff{
Feigin, B. and Frenkel E.: Quantum\ $ W$-algebras and elliptic algebras.
Preprint (1995) (q-alg/9508009)},\
\ref\wya{Shiraishi, J., Kubo, H., Awata,
 H. and Odake, S.: Quantum \ $W_N$\ algebras
and Macdonald Polynomials.
Preprint YITP/U-95-34, DPSU-95-9,
UT-718 (1995) (q-alg/9508011)}.

Let us recall some basic facts on DVA
in notations close to  \ \lp .
The algebra depends on two real parameters
\ $\xi>0$\ and \ $ 0<x<1$.
\foot{ J. Shiraishi, e.a.\ \yap\ use the notations
$ p=x^{-2}, \ q=x^{-2(\xi+1)},\ \ t=q p^{-1}=x^{-2\xi} \ . $}
Notice  that the parameters are related to  the constants
of interaction of the  XYZ chain\ \xyz\  as  follows:
\eqn\pare{\eqalign{&\frac{J_x-J_y}{J_x+J_y}=
\prod_{n=0}^{\infty}\frac{(1-x^{\xi+(\xi+1) n})^2(1-x^{1+(\xi+1) n})^2}
{(1+x^{\xi+(\xi+1) n})^2(1+x^{1+(\xi+1) n})^2 }\ ,\cr
&\frac{2J_z}{J_x+J_y}=
\frac{x^{\xi}}{4}\ \prod_{n=0}^{\infty}\frac{(1+x^{(1+\xi) n})^4
(1-x^{2\xi+(\xi+1) (2 n+1)})(1-x^{-2\xi+(\xi+1) (2 n+1)})}
{(1-x^{ (\xi+1)(2 n+1)})^2
(1+x^{\xi+(\xi+1) n})^2(1+x^{1+(\xi+1) n})^2}\ .}}
It is convenient to introduce the set of oscillators\ $\lambda_m$\
which satisfy the commutation relations:
\eqn\gs{[\lambda_m, \lambda_n]=\frac{1}{m}\ \frac{(x^{\xi m}-x^{-\xi m})
(x^{(\xi+1) m}-x^{-(\xi+1) m})}{x^m+x^{-m}}\ . }
Such  oscillators are  "self dual"   with respect to
the transform\
$\xi\to-\xi-1$.
\foot{The operators \ $\lambda_m$\   are connected with the
oscillators \ $b_m, b'_m$ \ from \lp\  in such a way:
$$b_m'\  (x^{(\xi+1)m}-x^{-(\xi+1)m})=b_m
(x^{\xi m}-x^{-\xi m})= m\ \lambda_m\ .$$}
We need  to
define also  the " zero mode" operators\ $ P,Q $, which  commute
with \ $\lambda_m$\ and satisfy the relation:
\eqn\ko{[Q,P]=i\ .}
Central objects in the bosonization procedure are
"screening charges"\
\ref\fei{Feigin, B.L.,  Fuchs, D.B.: Representations
of the Virasoro algebra. In: Topology, Proceedings, Leningrad 1982.
Faddeev, L.D., Mal'chev, A.A. (eds.). Lecture Notes in Mathematics,
vol. {\bf 1060.} Berlin Heidelberg, New York: Springer (1984)
},\
\ref\df{ Dotsenko,
 Vl. S. and Fateev, V. A.: Conformal algebra and
multi-point correlation functions in 2d statistical models. Nucl. Phys.
B{\bf 240}
\ [FS{\bf 12}], 312 (1984) \semi Dotsenko, Vl. S. and  Fateev, V. A.:
Four-point
correlation functions and the operator algebra in 2d conformal invariant
theories with
central charge $c\le1$. Nucl. Phys. B{\bf 251}\ [FS{\bf 13}] 691
(1985)},\ \ref\fel{Felder, G.: BRST approach to minimal models.
Nucl. Phys. B{\bf 317}, 215-236 (1989)}.
In terms of the  oscillators\ \gs ,\ \ko\ the "screening currents" can
be written in the form\ \lp:
\eqn\scr{\eqalign{&
I_+(z)=e^{-i \sqrt{\frac{2(\xi+1)}{\xi} } Q}
 z^{-\sqrt{\frac{2(\xi+1)}{\xi}} P+
\frac{1}{\xi}}\
{\rm :\exp}\Biggl(-\sum_{m\not= 0}\  \frac{x^m+x^{-m}}
{x^{\xi m}-x^{-\xi m}}\ \lambda_m\  z^{-m}\Biggr):\ ,\cr
&I_-(z)=
e^{i \sqrt{\frac{2\xi}{(\xi+1)}}Q} z^{\sqrt{\frac{2\xi}{(\xi+1)}} P
-\frac{1}{\xi+1}}\
{\rm :\exp}\Biggl(\sum_{m\not= 0}\  \frac{x^m+x^{-m}}
{x^{(\xi+1) m}-x^{-(\xi+1) m}}\ \lambda_m\ z^{-m} \Biggr):\ .
} }
Now introduce the field
\eqn\hy{\Lambda(z)=x^{\sqrt{2\xi (\xi+1)}P}\  :{\rm \exp}
\Biggl(-\sum_{m\not= 0} \lambda_m z^{-m}\Biggr):\ ,}
then a current of DVA
reads\ \yap ,\ \fre:
\eqn\jd{T(z)=\Lambda(z x^{-1})+\Lambda^{-1}(z x)\ .}
It commutes with the "screening charges"
\eqn\qom{[X^l_{\pm}, T(z)]=0\ ,}
where
$$X^l_{\pm}=
\int_{C_1}...\int_{C_l} \frac{d z_1}{2\pi i}...\frac{d z_l}{2\pi i}\
I_{\pm}(z_1)...I_{\pm}(z_l)\ ,$$
if the  integration contours
are  chosen according to   Felder' s
prescription\ \fel ,\ \lp  .
The field \  $T(z)$\ generates the algebra
with the   basic relation\ \yap :
\eqn\bas{\eqalign
{f(\zeta z^{-1})\  T(z)& T(\zeta)-f(z\zeta^{-1})\  T(\zeta) T(z)=\cr
& 2\pi \ \frac{(x^{\xi }-x^{-\xi })
(x^{(\xi+1) }-x^{-(\xi+1) })}
{x-x^{-1}}\ \big(\ \delta(\zeta   z^{-1} x^{-2})-
\delta(\zeta  z^{-1}  x^2)\ \big)\  ,}}
here
$$f(z)=
(1-z)^{-1}\ \prod_{n=0}^{\infty} \frac{(1-z x^{2(\xi+1)+4 n})
(1-z x^{-2\xi+4 n})}{(1-z x^{2(\xi+1)+2( 2 n+1)})
(1-z x^{-2\xi+2( 2 n+1)})} $$
and
$$\delta(z)=\frac{1}{2 \pi }\sum_{m=-\infty}^{+\infty} z^m\ .$$
This relation  can be equivalently written in terms of the modes\ $T_m$:
\ $T(z)=\sum_{m=-\infty}^{+\infty} T_m z^{-m}$
\eqn\vir{\sum_{l=0}^{+\infty}
f_l\  (T_{n-l} T_{m+l}-T_{m-l} T_{n+l})=
\frac{(x^{\xi }-x^{-\xi })
(x^{(\xi+1) }-x^{-(\xi+1) })}{x-x^{-1}}
(x^{2n}- x^{-2n}) \ \delta_{n+m,0}\ ,}
where \ $f_l$\ :
$$f(z)=\sum_{l=0}^{+\infty} f_l\   .$$
If we fix\ $z$ and consider the
limit \ $x\to1$\ ,
then
\eqn\dg{T(z)=
2+\xi (\xi+1)\ (x-x^{-1})^2\  \Biggl(   \sum_{m=-\infty}^{+\infty}
L_m z^{-m}+
\frac{1}{4 \xi (\xi+1)}\Biggr)+O((x-x^{-1})^4)\ .}
One can check that the
defining relations\ \vir\  give us the  Virasoro algebra
commutators  for the  modes \ $L_m$ and the corresponding
central charge is equal to
$$c=1-\frac{6}{\xi (\xi+1)}\ .$$

To explain the physical meaning of the field
\ $T(z)\ $ for the  XYZ model  let us introduce the
following parameterization
\eqn\para{ x^2=e^{-\pi \epsilon} , z=e^{- i\epsilon \beta}\ .}
After a little algebra, \ \bas\
can be rewritten in the form:
\eqn\ju{ T(\beta') T(\beta)=
S_{\epsilon}(\beta'-\beta)\   T(\beta) T(\beta')
+ C_{\epsilon}
\bigl(\  \delta(\beta'-\beta-i \pi)+\delta(\beta'-\beta+i \pi)
\ \bigr)\ ,}
where
\eqn\sm{S_{\epsilon}(\beta)=
\frac{\theta_1(i\frac{\beta}{2 \pi}-\frac{\xi}{2})\
\theta_2(i\frac{\beta}{2 \pi}+\frac{\xi}{2})}
{\theta_2(i\frac{\beta}{2 \pi}-\frac{\xi}{2})\
\theta_1(i\frac{\beta}{2 \pi}+\frac{\xi}{2})}. \ }
Here the functions \ $\theta_{1,2}(u)$\ are
the conventional  theta functions
with the
modular parameter $$\tau= i \epsilon^{-1}$$  and
the constant\ $ C_{\epsilon}$\ reads explicitly
\eqn\gl{C_{\epsilon}=\frac{2\pi}{\epsilon}
\prod_{n=0}^{\infty}\frac{(1-x^{2(\xi+1)+4n})(1-x^{-2\xi+4 n})}
{(1-x^{2\xi+4(n+1)})(1-x^{-2(\xi+1)+4 (n+1)})}\ .}
It is useful to
consider the limit\ $\epsilon\to 0$
in the commutation relation\ \ju\
. Now we will fix  \ $\beta
$. So  this limit differs from the "conformal" one,
when we fixed \ $ z$.
{}From the physical point of view this limit corresponds
to the  scaling limit of the infinite  XYZ-chain\ \xyz , when
it can be described by relativistic quantum field theory
with the Lagrangian \ \ref\lus{Luther, A.: Eigenvalue
spectrum of interacting massive fermions in one dimension.
Phys. Rev. {\bf B14}, 5, 2153-2159 (1976)}
\eqn\sg{L=\int_{-\infty}^{+\infty} d x\Biggl(\frac{1}{2}\
(\partial _{\mu} \phi)^2+\frac{m_0^2}{b^2} \cos(b\phi)\Biggr)\ }
where
$$b^2=8 \pi\  \frac{\xi}{\xi+1}\  .$$
The  "bare" mass\ $m_0$\  and the  physical mass scale
\ $ m_{phys}$\   in the Sine-Gordon model are connected with the
parameters of the initial  XYZ chain as
\eqn\ma{ m_0\sim \frac{ J_x-J_y}{J_x}\ ,\ \
 m_{phys}
\sim \Biggl(\frac{J_x-J_y}{J_x}\Biggr)^{\xi+1}
\sim e^{-\frac{\pi}{\epsilon}} \ .}
It is easy to see that in the scaling  limit the operator
\ $T(\beta)$ will generate the simple
Zamolodchikov - Faddeev  (ZF) algebra
\eqn\fr{T(\beta') T(\beta)=S(\beta'-\beta)\   T(\beta) T(\beta'),}
where both \ $\beta,\beta'$\ are real.
The function
\eqn\s{S(\beta)=\frac{\sinh\beta +
i\sin(\pi \xi)}{\sinh\beta - i\sin(\pi \xi)} }
coincides with the  two-particle S-matrix of the  basic
scalar particle in the Sine-Gordon model\ \ref\ver{
Vergeles, S. and Gryanik., V.: Two-dimensional
Quantum Field Theories having exact solutions.
Yadern. Fiz. {\bf 23}, 1324-1334 (1976)(in
Russian) }.
Moreover,  as it follows from\ \ju ,
the singular part of the operator product\ $T(\beta')T(\beta)$,
being considered as a function
of the complex variable\ $\beta'$ for real
\ $\beta$\ in the upper half plane\ $\Im m\ \beta'\geq 0$\ ,  contains
a  simple   pole at the point \ $\beta'=\beta+i \pi$\  with
a numerical residue.
This operator product  condition
together with the commutation relation\ \fr\
are the basic properties of ZF algebra
acting  in
the space of angular quantization\ \ref\lk{
Lukyanov, S.: Free field representation for Massive
Integrable Models. Commun. Math. Phys. {\bf 167}, 1, 183-226 (1995)},\
\ref\js{
Lukyanov, S.: Correlators of the Jost Functions in
the Sine-Gordon Model. Phys. Lett. {\bf B325}, 409-417 (1994)}.
In such a way we can identify
\ $T(\beta)$\   with the ZF operator of the  basic scalar particle.
Note that such  particle exists only for\  $0<\xi<1$,  so
this interpretation is valid only within this restriction.

Now let us go back to the  relation\ \ju\  and interpret it as
the  ZF algebra for the  basic scalar
excitation  in  the XYZ model \ \xyz
\ with\ $J_z>0$ .
The function\ \sm\  is
a two-particle   S-matrix of this excitation.
Having at hand the  oscillator representation\ \hy ,\  \jd\  for
the ZF operator of the
basic scalar  particle it is easy to get  the  form-factors
of local operators\ \jde ,\  \lk ,\ \js .
I hope  to return to this point
in future publications.

\listrefs

\end